**********************************************************************
**********************************************************************
\documentstyle[12pt]{article}
\begin{document}
\baselineskip 24pt
\hfuzz=1pt
\setlength{\textheight}{8.5in}
\setlength{\topmargin}{0in}
\begin{center}
\LARGE {\bf  Does Clauser-Horne-Shimony-Holt Correlation
or Freedman-Clauser
Correlation lead to the
largest violation of Bell's Inequality?
} \\  \vspace{.75in}
\large {M. Ardehali}\footnote[1]
{email address:ardehali@mel.cl.nec.co.jp}
\\ \vspace{.5in}
Research Laboratories,
NEC Corporation,\\
Sagamihara,
Kanagawa 229
Japan

\end{center}
\vspace{.40in}

\begin{abstract}

An inequality is deduced from
Einstein's locality and a supplementary assumption.
This inequality defines an experiment which can actually be 
performed with present technology to test local realism.
Quantum mechanics violate this inequality
a factor of 1.5.
In contrast, quantum mechanics
violates previous inequalities (for example,
Clauser-Horne-Shimony-Holt inequality of $1969$,
Freedman-Clauser inequality of $1972$,
Clauser-Horne inequality of $1974$)
by a factor of $\sqrt 2$.
Thus the magnitude of violation of the inequality derived in this
paper is
approximately $20.7\%$ larger than the magnitude of violation of
previous inequalities.
This result can be particularly important for the
experimental test of locality.
\end {abstract}
\pagebreak

The Copenhagen interpretation of quantum mechanics
is based on the fundamental assumption that the wave
function, with its statistical interpretation, provides
a complete description of physical reality.
This assumption
has been the object of severe criticism, most notably by
Einstein, who always maintained that the wave function should
be supplemented with additional ``hidden variables'' such that
these variables together with the wave function precisely determine
the results of individual experiments. In 1965, Bell
\cite{1} showed that the premises of locality and
realism, as postulated by
Einstein, Podolsky, and Rosen (EPR) \cite{2}, imply some
constrains on the statistics of two spatially separated
particles. These constrains, which are collectively known as
Bell inequalities, are sometime grossly violated by quantum
mechanics. Bell's theorem therefore is a proof
that all realistic interpretation
of quantum mechanics must be non-local.

Bell's original argument, however,
can not be experimentally tested
because it relies on perfect correlation of the spin of the two
particles \cite {3}. Faced with this problem,
Clauser-Horne-Shimony-Holt (CHSH) \cite{4},
Freedman-Clauser (FC) \cite{5}, and Clauser-Horne (CH) \cite{6}
derived correlation
inequalities for
systems which do not achieve $100\%$ correlation,
but which do achieve a necessary minimum correlation. 
Quantum mechanics
violates these inequalities
by as much as $\sqrt 2$
(the factor $\sqrt 2$ can be achieved only if rotational 
invariance is assumed, see Eq. $(4')$ of \cite{6}.
For a detailed discussion of CHSH, FC and CH inequalities, see the 
review article by Clauser and Shimony \cite{7}, especially
inequalities 5.3-5.7).
An experiment based on CHSH, or FC,
or CH inequality utilizes one-channel
polarizers in which the dichotomic choice is between the detection of
the photon and its lack of detection. A better experiment is
one in which
a truly binary choice
is made between the ordinary and the extraordinary rays.
In 1971, Bell \cite{8}, and later others [9-11],
derived correlation inequalities in which two-channel
polarizers
are used to test locality. Quantum mechanical probabilities
violate these 
inequalities also by a factor of $\sqrt 2$.
In this paper, we derive a correlation inequality
for two-channel polarizer systems and we show that 
quantum mechanics violates this inequality 
by a factor of $1.5$. Thus the magnitude of
violation of the inequality derived in this paper
is approximately $20.7\%$ larger than
the magnitude of violation of previous
inequalities of [4-11]. This result can be of considerable 
importance for the experimental test of local realism.

We start by considering the Bohm's \cite{12} version of EPR experiment
in which an unstable source emits pairs of photons in a singlet
state $\mid\!\!\Phi\rangle$. The source is viewed by two
apparatuses.
The first (second) apparatus consists of a polarizer
$P_1 \left(P_2 \right)$
set at angle $\mbox{\boldmath $a$} \left(
\mbox{\boldmath $b$} \right)$,
and two detectors
$D_{1}^{\,\pm} \left (D_{2}^{\,\pm} \right)$
put along the ordinary and the extraordinary beams.
During a period of time $T$, the source emits, say, $N$ pairs of
photons. Let $N^{\,\pm\,\pm}\left(\mbox{\boldmath $a,b$}\right)$
be the number of simultaneous counts
from detectors $D_{1}^{\pm}$ and $D_{2}^{\pm}$,
$N^{\,\pm}\left(\mbox{\boldmath $a$}\right)$
the number of counts from detectors
$D_1^\pm$, and
$N^{\,\pm}\left(\mbox{\boldmath $b$}\right)$
the number of counts from detectors
$D_2^\pm$.
If the time $T$ is sufficiently
long, then the ensemble probabilities
$p^{\;\pm\;\pm}\left(\mbox{\boldmath $a,b$}\right)$ are defined as
\begin{eqnarray}{\nonumber}
p^{\;\pm\;\pm} \left(\mbox{\boldmath $a,b$} \right)&=&
\frac{N^{\;\pm\;\pm} \left(\mbox{\boldmath $a,b$} \right)}{N}, \\
\nonumber
p^{\;\pm}(\mbox{\boldmath $a$})&=&
\frac{N^{\;\pm}(\mbox{\boldmath $a$})}{N}, \\ 
p^{\;\pm}(\mbox{\boldmath $b$})&=&
\frac{N^{\;\pm}(\mbox{\boldmath $b$})}{N}.
\end{eqnarray}

\noindent We consider a particular pair of photons and specify its
state with a parameter $\lambda$. Following Bell, we do not 
impose any restriction on the complexity of $\lambda$. 
``It is
a matter of indifference
in the following whether $\lambda$ denotes a single variable or
a set, or even a set of functions, and whether the variables are 
discrete or continuous.'' \cite{1}

The ensemble probabilities
in Eq. $(1)$ are defined as
\begin{eqnarray} {\nonumber}
p^{\:\pm\:\pm}(\mbox{\boldmath $a,b$}) &=&
\int p\,(\lambda)\, p^{\;\pm}(\mbox{\boldmath 
$a$} \mid \lambda) \, p^{\;\pm}(\mbox{\boldmath $b$}
\mid \lambda,\mbox{\boldmath $a$}), \\ \nonumber
p^{\:\pm}(\mbox{\boldmath $a$}) &=&
\int p \, (\lambda) \, p^{\;\pm}(\mbox{\boldmath 
$a$} \mid \lambda), \\ 
p^{\:\pm}(\mbox{\boldmath $b$}) &=&
\int p \, (\lambda)\, p^{\;\pm}(\mbox{\boldmath 
$b$} \mid \lambda).
\end{eqnarray}
Equations (2) may be stated in physical terms: The ensemble
probability for detection of photons by
detectors $D^{\;\pm}_{\; 1}$ and $D^{\;\pm}_{\;2}$
[that is $p^{\;\pm\;\pm}(\mbox{\boldmath $a,b$})$]
is equal to the sum or integral of the probability
that the emission is
in the state $\lambda$ [that is $p(\lambda)$], times the conditional
probability that if the emission is in the state $\lambda$,
then a count is triggered by the first detector $D^{\;\pm}_{1}$
[that is $p^{\;\pm}(\mbox{\boldmath $a$}
\mid \lambda)$],
times the conditional probability that 
if the emission is in the state 
$\lambda$ and if the first polarizer is set along axis $\boldmath a$,
then a count is triggered from the second detector $D^{\;\pm}_{2}$
[that is $p^{\;\pm}(\mbox{\boldmath $b$}
\mid \lambda,\mbox{\boldmath $a$})$].
Similarly the ensemble probability for detection of photons by
detector $D^{\;\pm}_{\;1} \left(D^{\;\pm}_{\;2} \right )$
{\large [} that is $p^{\;\pm}(\mbox{\boldmath $a$}) \left
[p^{\;\pm}(\mbox{\boldmath $b$}) \right]$ {\large ]}
is equal to the sum or integral of the probability that the photon
is in the state $\lambda$ [that is $p(\lambda)$], times the
conditional probability that if the
photon is in the state $\lambda$,
then a count is triggered by
detector $D^{\;\pm}_{1} \left(D^{\;\pm}_{2} \right )$
{\large[} that is $p^{\;\pm}(\mbox{\boldmath $a$}
\mid \lambda) \left [p^{\;\pm}(\mbox{\boldmath $b$} \mid 
\lambda ) \right ]$ {\large]}.
Note that Eqs. $(1)$ and $(2)$ are quite general and follow
from the standard rules of probability theory.
No assumption has yet been made that is not satisfied 
by quantum mechanics.

Hereafter, we
focus our attention only on those theories that satisfy 
Einstein's criterion of locality,
``But on one supposition we should, in my
opinion absolutely hold fast: the real factual situation of the system
$S_2$ is independent of what is done with the
system $S_1$, which is spatially separated
from the former'' \cite {13}.
Einstein's criterion of locality can be translated into
the following mathematical equation:
\begin{eqnarray}
p^{\;\pm}(\mbox{\boldmath $b$} \mid \lambda,
\mbox{\boldmath $a$})=
p^{\;\pm}(\mbox{\boldmath $b$} \mid \lambda).
\end{eqnarray}
\noindent Equation $(3)$ is the hall mark of local realism. It
is the most general form of locality that accounts
for correlations subject only to the requirement that a count
from the second detector does not depend on
the orientation of the first polarizer. The assumption
of locality as postulated by Einstein, i.e., Eq. $(3)$, is
quite natural since the two photons are spatially separated so that
the orientation of the first polarizer should not influence the
measurement carried out on the second photon. Now substituting Eq.
(3) in Eq. (2), we obtain ensemble probabilities that 
satisfy Einstein's criterion of locality:
\begin{eqnarray} {\nonumber}
p^{\:\pm\:\pm}(\mbox{\boldmath $a,b$}) &=&
\int p\,(\lambda)\, p^{\;\pm}(\mbox{\boldmath
$a$} \mid \lambda) \, p^{\;\pm}(\mbox{\boldmath $b$}
\mid \lambda), \\ \nonumber
p^{\:\pm}(\mbox{\boldmath $a$}) &=&
\int p \, (\lambda) \, p^{\;\pm}(\mbox{\boldmath
$a$} \mid \lambda), \\
p^{\:\pm}(\mbox{\boldmath $b$}) &=&
\int p \, (\lambda)\, p^{\;\pm}(\mbox{\boldmath
$b$} \mid \lambda).
\end{eqnarray}

Before proceeding any further, it is useful to describe the
difference between Eq. (3) and CH's
criterion of locality. CH write their assumption of
locality as
\begin{eqnarray}
p^+ \left(\mbox{\boldmath $a, b$} , \lambda \right)=
p^+ \left(\mbox{\boldmath $a$} , \lambda \right)
p^+ \left(\mbox{\boldmath $b$}, \lambda \right).
\end{eqnarray}
Apparently by $p^+ \left(\mbox{\boldmath $a, b$} , \lambda \right)$, 
they mean the conditional probability that if the
emission is in state $\lambda$, 
then simultaneous counts are triggered by detectors
$D^+_1$ and $D^+_2$.  However, what they call 
$p^+ \left(\mbox{\boldmath $a, b$} , \lambda \right)$ 
in probability theory is usually
written as
$p^+ \left(\mbox{\boldmath $a, b$} \mid\lambda \right)$ [note
that $p (x, y, z)$ is the joint
probability of $x, y$ and $z$, whereas $p (x, y \mid z)$ 
is the conditional probability that
if $z$ then $x$ and $y$]. Similarly by
$p^+ \left(\mbox{\boldmath $a$} , \lambda \right) {\large [}
p^+ \left(\mbox{\boldmath $b$}, \mid \lambda \right){\large]}$,
CH mean the conditional
probability that if the emission is in state $\lambda$,
then a count is triggered from the
detector $D^+_1 \left(D^+_2 \right)$.
Again what they call 
$p^+ \left(\mbox{\boldmath $a$} , \lambda \right) {\large [}
p^+ \left(\mbox{\boldmath $b$}, \lambda \right){\large]}$
in probability
theory is usually written as 
$p^+ \left(\mbox{\boldmath $a$} \mid \lambda \right) {\large [}
p^+ \left(\mbox{\boldmath $b$} \mid \lambda \right){\large]}$
(again note that $p(x, z)$ is the
joint probability of $x$ and $z$, whereas $p (x \mid z)$
is the conditional probability
that if $z$ then $x$).
Thus according to standard notation of probability theory,
CH criterion of locality may be written as
\begin{eqnarray} 
p^+ \left(\mbox{\boldmath $a, b$} \mid \lambda \right)=
p^+ \left(\mbox{\boldmath $a$} \mid \lambda \right)
p^+ \left(\mbox{\boldmath $b$} \mid \lambda \right).
\end{eqnarray}
Now according to Bayes' theorem,
\begin{eqnarray}
p^+ \left(\mbox{\boldmath $a, b$} \mid \lambda \right)=
p^+ \left(\mbox{\boldmath $a$} \mid \lambda \right)
p^+ \left(\mbox{\boldmath $b$} \mid
\lambda , \mbox{\boldmath $a$} \right).
\end{eqnarray}
Substituting Eq. (7) in Eq. (6), we obtain
\begin{eqnarray}
p^+ \left(\mbox{\boldmath $ b$}
\mid \lambda, \mbox{\boldmath $a$} \right)=
p^+ \left(\mbox{\boldmath $b$} \mid \lambda\right),
\end{eqnarray}
which for the ordinary equation is the same as Eq. (3).

Having clarified the difference between Eq. (3) and CH's criterion of 
locality, we now show that 
Eqs. $(4)$ lead to validity of an equality
that is sometimes grossly violated by
the quantum mechanical predictions in the case of real experiments.
First we need to prove the following algebraic theorem.

{\it Theorem:} Given ten non-negative real numbers
$x_{1}^{+}$, $x_{1}^{-}$, $x_{2}^{+}$, $x_{2}^{-}$,
$y_{1}^{+}$, $y_{1}^{-}$, $y_{2}^{+}$, $y_{2}^{-}$, $U$ and $V$
such that
$x_{1}^{+}, x_{1}^{-},
x_{2}^{+}, x_{2}^{-} \leq U$,
and
$y_{1}^{+}, y_{1}^{-},
y_{2}^{+}, y_{2}^{-} \leq V$,
then the following inequality always holds:
\begin{eqnarray}{\nonumber}
Z &=& x_{1}^{+}y_{1}^{+}
+x_{1}^{-}y_{1}^{-}
-x_{1}^{+}y_{1}^{-}
-x_{1}^{-}y_{1}^{+}
+x_{1}^{+}y_{2}^{+}
+x_{1}^{-}y_{2}^{-} \\ \nonumber
&-&x_{1}^{+}y_{2}^{-}
-x_{1}^{-}y_{2}^{+}
+x_{2}^{+}y_{1}^{+}
+x_{2}^{-}y_{1}^{-}
-x_{2}^{+}y_{1}^{-}
-x_{2}^{-}y_{1}^{+}
-2x_{2}^{+}y_{2}^{+} \\
&-&2x_{2}^{-}y_{2}^{-}
+Vx_{2}^{+}+Vx_{2}^{-}
+Uy_{2}^{+}+Uy_{2}^{-}+UV
\ge 0.
\end{eqnarray}
{\it Proof}:
Calling $A=y_{1}^{+}-y_{1}^{-}$, we write the function $Z$ as
\begin{eqnarray} {\nonumber}
Z&=&
x_{2}^{+}
\left(-2y_{2}^{+} + A + V \right )
+ x_{2}^{-}
\left(-2y_{2}^{-} - A + V \right ) \\
&+&\left( x_{1}^{+} - x_{1}^{-} \right )
\left(A + y_{2}^{+}-y_{2}^{-} \right )
+ U y_{2}^{+} +  U y_{2}^{-}
+UV.
\end{eqnarray}

\noindent We consider the following eight cases:
\\
(1) First assume

\vspace{0.3 cm}
$\left \{
\begin{array}{c}
-2y_{2}^{+} + A  + V \ge 0,\\
-2y_{2}^{-} - A + V \ge 0,\\
A + y_{2}^{+}-y_{2}^{-}\ge 0.
\end{array} \right.$
\vspace{0.4 cm}

\noindent The function $Z$ is minimized if
$x_{2}^{+}=0, x_{2}^{-}=0$, and
$ x_{1}^{+} - x_{1}^{-} =-U$. Thus
\begin{eqnarray} {\nonumber}
Z &\ge&
-U \left(A +y_{2}^{+} - y_{2}^{-} \right )
+U y_{2}^{+} +  U y_{2}^{-}
+UV \\
&=&U\left(-A + 2y_{2}^{-} + V\right ).
\end{eqnarray}
Since $V \ge A$ and $y_{2}^{-} \ge 0$, $Z \ge 0$.
\vspace{0.7 cm}
\\
(2) Next assume
$\left \{
\begin{array}{c}
-2y_{2}^{+} + A + V < 0,\\
-2y_{2}^{-} - A + V \ge 0,\\
A + y_{2}^{+}-y_{2}^{-}\ge 0.
\end{array} \right.$
\vspace{0.4 cm}

\noindent The function $Z$ is minimized if
$x_{2}^{+}=U, x_{2}^{-}=0$, and
$ x_{1}^{+} - x_{1}^{-} =-U$. Thus
\begin{eqnarray} {\nonumber}
Z &\ge&
U
\left(-2y_{2}^{+} + A + V \right ) -
U \left(A + y_{2}^{+}-y_{2}^{-} \right )
+ U y_{2}^{+} +  U y_{2}^{-}
+UV \\
&=&2U\left(V+y_{2}^{-}-y_{2}^{+}\right ).
\end{eqnarray}
Since $V \ge y_{2}^{+}$, and $y_{2}^{-} \ge 0$, $Z \ge 0$.
\vspace{0.7 cm}
\\
(3) Next assume
$\left \{
\begin{array}{c}
-2y_{2}^{+} + A + V \ge 0,\\
-2y_{2}^{-} - A + V < 0,\\
A + y_{2}^{+}-y_{2}^{-}\ge 0.
\end{array} \right.$
\vspace{0.4 cm}

\noindent The function $Z$ is minimized if
$x_{2}^{+}=0, x_{2}^{-}=U$, and
$ x_{1}^{+} - x_{1}^{-} =-U$. Thus
\begin{eqnarray} {\nonumber}
Z &\ge&
U
\left(-2y_{2}^{-} - A + V \right ) -
U \left(A + y_{2}^{+}-y_{2}^{-} \right )
+ U y_{2}^{+} +  U y_{2}^{-}
+UV \\
&=&2U\left (V - A \right).
\end{eqnarray}
Since $V \ge  A$, $Z \ge 0$.
\vspace{0.7 cm}
\\
(4) Next assume
$\left \{
\begin{array}{c}
-2y_{2}^{+} + A + V \ge 0,\\
-2y_{2}^{-} - A + V \ge 0,\\
A + y_{2}^{+}-y_{2}^{-} < 0.
\end{array} \right.$
\vspace{0.4 cm}

\noindent The function $Z$ is minimized if
$x_{2}^{+}=0, x_{2}^{-}=0$, and
$ x_{1}^{+} - x_{1}^{-} =U$. Thus
\begin{eqnarray} {\nonumber}
Z &\ge&
U \left(A + y_{2}^{+}-y_{2}^{-} \right )
+ U y_{2}^{+} +  U y_{2}^{-}
+UV \\
&=&U\left(A + 2y_{2}^{+} + V\right ).
\end{eqnarray}
Since $V \ge A$ and $y_{2}^{+} \ge 0$, $Z \ge 0$.
\vspace{0.7 cm}
\\
(5) Next assume
$\left \{
\begin{array}{c}
-2y_{2}^{+} + A + V < 0,\\
-2y_{2}^{-} - A + V < 0,\\
A + y_{2}^{+}-y_{2}^{-}\ge 0.
\end{array} \right.$
\vspace{0.4 cm}

\noindent The function $Z$ is minimized if
$x_{2}^{+}=U, x_{2}^{-}=U$, and
$ x_{1}^{+} - x_{1}^{-} =-U$. Thus
\begin{eqnarray} {\nonumber}
Z &\ge&
U
\left(-2y_{2}^{+} + A + V \right )
+U
\left(-2y_{2}^{-} - A + V \right )
-U \left(A + y_{2}^{+}-y_{2}^{-} \right )\\ \nonumber
&+& U y_{2}^{+} +  U y_{2}^{-}
+UV \\ 
&=&U\left(-2y_{2}^{+} -A +3V \right ).
\end{eqnarray}
Since $V \ge A$ and $V \ge y_{2}^{+}$, $Z \ge 0$.
\vspace{0.7 cm}
\\
(6) Next assume
$\left \{
\begin{array}{c}
-2y_{2}^{+} + A + V < 0,\\
-2y_{2}^{-} - A + V \ge 0,\\
A + y_{2}^{+}-y_{2}^{-} < 0.
\end{array} \right.$
\vspace{0.4 cm}

\noindent The function $Z$ is minimized if
$x_{2}^{+}=U, x_{2}^{-}=0$, and
$ x_{1}^{+} - x_{1}^{-} =U$. Thus
\begin{eqnarray} {\nonumber}
Z &\ge&
U\left(-2y_{2}^{+} + A + V \right )+
U \left(A + y_{2}^{+}-y_{2}^{-} \right )
+ U y_{2}^{+} +  U y_{2}^{-}
+UV \\ 
&=&2U\left(A + V \right ).
\end{eqnarray}
Since $V \ge A$, $Z \ge 0$.
\vspace{0.7 cm}
\\
(7) Next assume
$\left \{
\begin{array}{c}
-2y_{2}^{+} + A + V \ge 0,\\
-2y_{2}^{-} - A + V < 0,\\
A + y_{2}^{+}-y_{2}^{-} < 0.
\end{array} \right.$
\vspace{0.4 cm}

\noindent The function $Z$ is minimized if
$x_{2}^{+}=0, x_{2}^{-}=U$, and
$ x_{1}^{+} - x_{1}^{-} = U$. Thus
\begin{eqnarray} {\nonumber}
Z &\ge&
U
\left(-2y_{2}^{-} - A + V \right )
+U \left(A + y_{2}^{+}-y_{2}^{-} \right )
+ U y_{2}^{+} +  U y_{2}^{-}
+UV \\ 
&=&2U\left( y_{2}^{+}-y_{2}^{-} + V \right ).
\end{eqnarray}
Since $V \ge y_{2}^{-}$ and $y_{2}^{+} \ge 0$, $Z \ge 0$.
\vspace{0.7 cm}
\\
(8) Finally assume
$\left \{
\begin{array}{c}
-2y_{2}^{+} + A + V < 0,\\
-2y_{2}^{-} - A + V < 0,\\
A + y_{2}^{+}-y_{2}^{-} < 0.
\end{array} \right.$
\vspace{0.3 cm}

\noindent The function $Z$ is minimized if
$x_{2}^{+}=U, x_{2}^{-}=U$, and
$ x_{1}^{+} - x_{1}^{-} =U$. Thus
\begin{eqnarray} {\nonumber}
Z &\ge&
U
\left(-2y_{2}^{+} + A +V \right )
+U
\left(-2y_{2}^{-} - A + V \right )
+U
\left(A + y_{2}^{+}-y_{2}^{-} \right ) \\ \nonumber
&+& U y_{2}^{+} +  U y_{2}^{-}
+UV \\
&=& U\left(-2y_{2}^{-} + A + 3V \right ).
\end{eqnarray}
Since $V \ge A$ and $V \ge y_{2}^{-}$, $Z \ge 0$,
and the theorem is proved.

Now let $\mbox {\boldmath $a$ ($b$)}$
and $\mbox {\boldmath $a'$ ($b'$)}$
be two arbitrary orientation of the first
(second) polarizer, and let
\begin{eqnarray}{\nonumber}
x_{1}^{\pm}&=&p^{\;\pm}(\mbox{\boldmath $a$} \mid \lambda), \qquad
x_{2}^{\pm}=p^{\;\pm}(\mbox{\boldmath $a'$}|\lambda), \\ 
y_{1}^{\pm}&=&p^{\;\pm}(\mbox{\boldmath $b$}|\lambda), \qquad
y_{2}^{\pm}=p^{\;\pm}(\mbox{\boldmath $b'$}|\lambda).
\end{eqnarray}

\noindent Obviously for each value of $\lambda$, we have
\begin{eqnarray}{\nonumber}
p^{\;\pm}(\mbox{\boldmath $a$} \mid \lambda) \leq 1, \qquad
p^{\;\pm}(\mbox{\boldmath $a'$} \mid \lambda) \leq 1,\\ 
p^{\;\pm}(\mbox{\boldmath $b$} \mid \lambda) \leq 1, \qquad
p^{\;\pm}(\mbox{\boldmath $b'$} \mid \lambda) \leq 1.
\end{eqnarray}

\noindent Inequalities ($9$) and ($20$) yield
\begin{eqnarray} {\nonumber}
&&p^{+}(\mbox{\boldmath $a$} \mid \lambda)
p^{+}(\mbox{\boldmath $b$} \mid \lambda)
+p^{-}(\mbox{\boldmath $a$} \mid \lambda)
p^{-}(\mbox{\boldmath $b$} \mid \lambda)
-p^{+}(\mbox{\boldmath $a$} \mid \lambda) 
p^{-}(\mbox{\boldmath $b$} \mid \lambda)  \\ \nonumber
&&-p^{-}(\mbox{\boldmath $a$} \mid \lambda)  
p^{+}(\mbox{\boldmath $b$} \mid \lambda)+
p^{+}(\mbox{\boldmath $a$} \mid \lambda)
p^{+}(\mbox{\boldmath $b'$} \mid \lambda) 
+p^{-}(\mbox{\boldmath $a$} \mid \lambda)
p^{-}(\mbox{\boldmath $b'$} \mid \lambda) \\ \nonumber
&&-p^{+}(\mbox{\boldmath $a$} \mid \lambda)
p^{-}(\mbox{\boldmath $b'$} \mid \lambda) -
p^{-}(\mbox{\boldmath $a$} \mid \lambda) 
p^{+}(\mbox{\boldmath $b'$} \mid \lambda)+
p^{+}(\mbox{\boldmath $a'$} \mid \lambda)
p^{+}(\mbox{\boldmath $b$} \mid \lambda) \\ \nonumber
&&+p^{-}(\mbox{\boldmath $a'$} \mid \lambda)
p^{-}(\mbox{\boldmath $b$} \mid \lambda) -
p^{+}(\mbox{\boldmath $a'$} \mid \lambda) 
p^{-}(\mbox{\boldmath $b$} \mid \lambda)
-p^{-}(\mbox{\boldmath $a'$} \mid \lambda)
p^{+}(\mbox{\boldmath $b$} \mid \lambda) \\ \nonumber
&&-2p^{+}(\mbox{\boldmath $a'$} \mid \lambda) 
p^{+}(\mbox{\boldmath $b'$} \mid \lambda)-
2p^{-}(\mbox{\boldmath $a'$} \mid \lambda)
p^{-}(\mbox{\boldmath $b'$} \mid \lambda)+ 
p^{+}(\mbox{\boldmath $a'$} \mid \lambda) \\
&&+p^{-}(\mbox{\boldmath $a'$} \mid \lambda)
+p^{+}(\mbox{\boldmath $b'$} \mid \lambda)
+p^{-}(\mbox{\boldmath $b'$} \mid \lambda) \ge -1.
\end{eqnarray}

\noindent Multiplying both sides of $(21)$
by $p(\lambda)$, integrating over $\lambda$ and
using Eqs. $(4)$, we obtain
\begin{eqnarray} {\nonumber}
&&p^{+ +}(\mbox{\boldmath $a,\, b$}) 
+p^{- \,-}(\mbox{\boldmath $a,\, b$}) 
-p^{+ \,-}(\mbox{\boldmath $a,\, b$}) 
-p^{- \,+}(\mbox{\boldmath $a,\, b$})  
+p^{+ +}(\mbox{\boldmath $a,\, b'$})+ \\ \nonumber
&&p^{- \,-}(\mbox{\boldmath $a,\, b'$})
-p^{+ \,-}(\mbox{\boldmath $a,\, b'$})
-p^{- \,+}(\mbox{\boldmath $a,\, b'$}) 
+p^{+ +}(\mbox{\boldmath $a',\, b$})+ \\ \nonumber
&&p^{- \,-}(\mbox{\boldmath $a',\, b$}) 
-p^{+ \,-}(\mbox{\boldmath $a',\, b$}) 
-p^{- \,+}(\mbox{\boldmath $a',\, b$}) 
-2p^{+ +}(\mbox{\boldmath $a',\, b'$}) - \\
&&2p^{- \,-}(\mbox{\boldmath $a',\, b'$}) 
+p^{+}(\mbox{\boldmath $a'$}) 
+p^{-}(\mbox{\boldmath $a'$}) 
+p^{+}(\mbox{\boldmath $b'$}) 
+p^{-}(\mbox{\boldmath $b'$})
\geq -1.
\end{eqnarray}
All local realistic theories must satisfy inequality $(22)$.

In the atomic cascade experiments, an atom emits two photons in 
a cascade from state $J=1$ to $J=0$. Since the pair of photons
have zero angular momentum, they propagate in the form of spherical
wave. Thus the probability $p \left(\mbox{\boldmath $d_1$},
\mbox{\boldmath $d_2$} \right)$ 
of both photons being simultaneously detected
by two detectors in the directions $\mbox{\boldmath $d_1$}$ and
$\mbox{\boldmath $d_2$}$ is 
\begin{eqnarray}
p \left(\mbox{\boldmath $d_1,\,d_2$} \right)=
\eta^2 \left ({\frac{\displaystyle \Omega}
{\displaystyle 4\pi}}\right) ^2
g \left (\theta,\phi \right ),
\end{eqnarray}
where $\cos \theta=\mbox{\boldmath $d_1. d_1$}$,
$\eta$ is the quantum efficiency of the detectors, 
$\Omega$ is the solid angle of the detector, 
and angle $\phi$ is related to $\Omega$ by
\begin{eqnarray}
\Omega=2 \pi \left (1-\cos \phi \right).
\end{eqnarray}
Finally the function 
$g \left (\theta,\phi \right )$ is the angular correlation function
and in the special cases is given by
\begin{eqnarray} {\nonumber}
g \left (\theta, 0 \right ) =
\frac{3}{4} \left (1 + \cos^2 \theta \right),\\ 
g \left (\pi, \phi \right ) = 1+
\frac{1}{8} \cos^2 \phi \left (1 + \cos \phi \right)^2.
\end{eqnarray}
If we insert polarizers in front of the detectors, then the
quantum mechanical predictions for joint detection probabilities are
\begin{eqnarray} {\nonumber}
p^{+} \left ( \mbox{\boldmath $a$} \right )=
p^{-} \left ( \mbox{\boldmath $a$} \right )=
\eta \left ({\frac{\displaystyle \Omega}{\displaystyle 8 \pi}}
\right), \qquad
p^{+} \left ( \mbox{\boldmath $b$} \right )=
p^{-} \left ( \mbox{\boldmath $b$} \right )=
\eta \left ({\frac{\displaystyle \Omega}{\displaystyle 8 \pi}}
\right), \\ \nonumber
p^{+ +} \left ( \mbox{\boldmath $a,\, b$} \right )=
p^{- -} \left ( \mbox{\boldmath $a,\, b$} \right )=
\eta^2 \left ({\frac{\displaystyle \Omega}{\displaystyle 8 \pi}}
\right)^2
g \left (\theta,\phi \right )
\left[1+F \left (\theta,\phi \right ) 
cos 2 \left ( \mbox{\boldmath $a- b$} \right ) \right ], \\ 
p^{+ -} \left ( \mbox{\boldmath $a,\, b$} \right )=
p^{- +} \left ( \mbox{\boldmath $a,\, b$} \right )=
\eta^2 {\frac{\displaystyle \Omega}{\displaystyle 8 \pi}}^2
g \left (\theta,\phi \right )
\left[1-F \left (\theta,\phi \right ) 
\cos 2\left ( \mbox{\boldmath $a- b$} \right ) \right].
\end {eqnarray}

In experiments which are feasible with present technology \cite {14},
because $\Omega \ll 4 \pi$,
only a very small fraction of photons are detected. Thus inequality
$(22)$ can not be used to test the violation of Bell's 
inequality \cite{15}. We now
state a supplementary assumption, and we
show that this assumption is sufficient to make these
experiments (where $\Omega \ll 4 \pi$)
applicable as a test of local theories.
The supplementary assumption is:
For every emission $\lambda$, the detection probability 
by detector $D^{+}$ or $D^-$
is {\it less than or equal} to the sum  of detection probabilities
by detectors $D^{+}$ and $D^-$ when the polarizer
is set along any {\it arbitrary} axis.
If we let $\mbox{\boldmath $r$}$ be an arbitrary direction of the 
first polarizer and
$\mbox{\boldmath $s$}$ be an arbitrary direction of the 
second polarizer, then the above supplementary assumption
may be translated into the 
following inequalities
\begin{eqnarray} {\nonumber}
p^{\;+}(\mbox{\boldmath $a$} \mid \lambda) \leq 
p^{\;+}(\mbox{\boldmath $r$} \mid \lambda)+
p^{\;-}(\mbox{\boldmath $r$} \mid \lambda), \qquad
p^{\;-}(\mbox{\boldmath $a$} \mid \lambda) \leq 
p^{\;+}(\mbox{\boldmath $r$} \mid \lambda)+
p^{\;-}(\mbox{\boldmath $r$} \mid \lambda), \\ \nonumber
p^{\;+}(\mbox{\boldmath $a'$} \mid \lambda) \leq 
p^{\;+}(\mbox{\boldmath $r$} \mid \lambda)+
p^{\;-}(\mbox{\boldmath $r$} \mid \lambda), \qquad
p^{\;-}(\mbox{\boldmath $a'$} \mid \lambda) \leq 
p^{\;+}(\mbox{\boldmath $r$} \mid \lambda)+
p^{\;-}(\mbox{\boldmath $r$} \mid \lambda), \\ \nonumber
p^{\;+}(\mbox{\boldmath $b$} \mid \lambda) \leq 
p^{\;+}(\mbox{\boldmath $s$} \mid \lambda)+
p^{\;-}(\mbox{\boldmath $s$} \mid \lambda), \qquad
p^{\;-}(\mbox{\boldmath $b$} \mid \lambda) \leq 
p^{\;+}(\mbox{\boldmath $s$} \mid \lambda)+
p^{\;-}(\mbox{\boldmath $s$} \mid \lambda), \\  \nonumber
p^{\;+}(\mbox{\boldmath $b'$} \mid \lambda) \leq 
p^{\;+}(\mbox{\boldmath $s$} \mid \lambda)+
p^{\;-}(\mbox{\boldmath $s$} \mid \lambda), \qquad
p^{\;-}(\mbox{\boldmath $b'$} \mid \lambda) \leq 
p^{\;+}(\mbox{\boldmath $s$} \mid \lambda)+
p^{\;-}(\mbox{\boldmath $s$} \mid \lambda).
\\
\end{eqnarray}
\noindent Now using relations $(4), (9), (27)$, and applying
the same argument that
led to inequality $(22)$, we obtain the following inequality
\begin{eqnarray} {\nonumber}
&&p^{+ \,+}(\mbox{\boldmath $a,\, b$})
+p^{- \,-}(\mbox{\boldmath $a,\, b$})
-p^{+ \,-}(\mbox{\boldmath $a,\, b$})
-p^{- \,+}(\mbox{\boldmath $a,\, b$}) 
+p^{+ +}(\mbox{\boldmath $a,\, b'$})   
+p^{- \,-}(\mbox{\boldmath $a,\, b'$})   \\  \nonumber
&&-p^{+ \,-}(\mbox{\boldmath $a,\, b'$}) 
-p^{- \,+}(\mbox{\boldmath $a,\, b'$}) 
+p^{+ \,+}(\mbox{\boldmath $a',\, b$}) 
+p^{- \,-}(\mbox{\boldmath $a',\, b$})
-p^{+ \,-}(\mbox{\boldmath $a',\, b$}) \\  \nonumber
&&-p^{- \,+}(\mbox{\boldmath $a',\, b$}) 
-2p^{+ +}(\mbox{\boldmath $a',\, b'$})
-2p^{- \,-}(\mbox{\boldmath $a',\, b'$})
+p^{+\,+}(\mbox{\boldmath $a',s$})
+p^{+\,-}(\mbox{\boldmath $a',s$}) \\  \nonumber
&&+p^{-\,+}(\mbox{\boldmath $a',s$}) 
+p^{-\,-}(\mbox{\boldmath $a',s$}) 
+p^{+\,+}(\mbox{\boldmath $r,b'$})
+p^{+\,-}(\mbox{\boldmath $r,b'$})
+p^{-\, +}(\mbox{\boldmath $r,b'$})\\  \nonumber
&&+p^{-\, -}(\mbox{\boldmath $r,b'$}) 
\left[p^{+\,+}(\mbox{\boldmath $s,r$}) 
+p^{+\,-}(\mbox{\boldmath $s,r$})
+p^{-\,+}(\mbox{\boldmath $s,r$})
+p^{-\,-}(\mbox{\boldmath $s,r$}) \right ]^{-1} \geq -1.
\\
\end{eqnarray}
\noindent Note that in the above inequality the
the number of emission from the source
(something which can not be measured experimentally)
is eliminated from the ratio. In fact, in
terms of measured
detection numbers (something which can be measured experimentally), the
above inequality may be written as
\begin{eqnarray} {\nonumber}
&&N^{+ \,+}(\mbox{\boldmath $a,\, b$})
+N^{- \,-}(\mbox{\boldmath $a,\, b$})
-N^{+ \,-}(\mbox{\boldmath $a,\, b$})
-N^{- \,+}(\mbox{\boldmath $a,\, b$}) 
+N^{+ +}(\mbox{\boldmath $a,\, b'$}) 
+N^{- \,-}(\mbox{\boldmath $a,\, b'$}) \\  \nonumber
&&-N^{+ \,-}(\mbox{\boldmath $a,\, b'$}) 
-N^{- \,+}(\mbox{\boldmath $a,\, b'$}) 
+N^{+ \,+}(\mbox{\boldmath $a',\, b$}) 
+N^{- \,-}(\mbox{\boldmath $a',\, b$})
-N^{+ \,-}(\mbox{\boldmath $a',\, b$})  \\  \nonumber
&&-N^{- \,+}(\mbox{\boldmath $a',\, b$}) 
-2N^{+ +}(\mbox{\boldmath $a',\, b'$})  
-2N^{- \,-}(\mbox{\boldmath $a',\, b'$})
+N^{+\,+}(\mbox{\boldmath $a',s$})
+N^{+\,-}(\mbox{\boldmath $a',s$}) \\  \nonumber
&&+N^{-\,+}(\mbox{\boldmath $a',s$}) 
+N^{-\,-}(\mbox{\boldmath $a',s$})
+N^{+\,+}(\mbox{\boldmath $r,b'$})
+N^{+\,-}(\mbox{\boldmath $r,b'$})
+N^{-\, +}(\mbox{\boldmath $r,b'$}) \\  \nonumber
&&+N^{-\, -}(\mbox{\boldmath $r,b'$}) 
\left[N^{+\,+}(\mbox{\boldmath $s,r$}) 
+N^{+\,-}(\mbox{\boldmath $s,r$})
+N^{-\,+}(\mbox{\boldmath $s,r$})
+N^{-\,-}(\mbox{\boldmath $s,r$}) \right ]^{-1} \geq -1.
\\
\end{eqnarray}
Inequality $(29)$ contains only double-detection
probabilities. Quatum mechanics violates
this inequality
in case of real experiments where the solid angle covered 
by the aperture of the apparatus, $\Omega$, is  much less than
$4 \pi$. In particular
the magnitude of violation
is maximized if the following set of orientations are chosen:
$\mbox{\boldmath $(a,\, b)$}=\mbox{\boldmath $(a,\, b')$}=
\mbox{\boldmath $(a',\, b)$}=120^\circ $, and 
$\mbox{\boldmath $(a',\, b')$}=0^\circ $.
Using the quantum mechanical probabilities
(i.e., Eqs. $(26)$),
inequality $(28)$ [or $(29)$] becomes
$-1.5 \geq -1$, 
which is certainly impossible.

In summary, we have demonstrated that
the conjunction of Einstein's locality
[Eq. $(3)$] with a supplementary assumption [inequality $(27)$]
leads to validity of inequality $(29)$
that is sometimes grossly violated
by quantum mechanics.
Inequality $(29)$, which may be called weak inequality [3,16,17],
defines an experiment which can actually
be performed with present technology and which does not require
the number of emissions $N$. Quantum mechanics violates this 
inequality by a factor of 1.5, whereas it violates the
previous inequalities (CHSH inequality
of $1969$ \cite{4}, FC inequality
of $1972$ \cite{5}, CH inequality of $1974$ \cite{6}, 
Bell's inequality of $1971$ \cite{8}, and inequalities of [9-11])
by a factor of $\sqrt 2$.
Thus the magnitude of violation of the inequality derived in this
paper is
approximately $20.7\%$ larger than the magnitude of violation of
the previous inequalities.
The larger violation of Bell's 
inequality can be of considerable importance for the experimental
test of locality.

\pagebreak

\end{document}